\documentclass[pre,twocolumn,showpacs,superscriptaddress]{revtex4-1}
\usepackage{graphicx}
\usepackage{amsmath}
\usepackage{amssymb}
\usepackage{amsbsy}
\usepackage{hyperref}
\usepackage[usenames]{color}

\begin{document}
\title{Stochastic resonance in the two-dimensional $q$-state clock models}
\author{Hye Jin Park}
\affiliation{Department of Physics, Sungkyunkwan University, Suwon 440-746, Korea}
\author{Seung Ki Baek}
\email[Corresponding author: ]{seungki@pknu.ac.kr}
\affiliation{Department of Physics, Pukyong National University, Busan 608-737,
Korea}
\author{Beom Jun Kim}
\email[Corresponding author: ]{beomjun@skku.edu}
\affiliation{Department of Physics, Sungkyunkwan University, Suwon 440-746, Korea}

\begin{abstract}
We numerically study stochastic resonance in the two-dimensional $q$-state
clock models from $q=2$ to $7$ under a weak oscillating magnetic field. As
in the mean-field case, we observe double resonance peaks, but the detailed
response strongly depends on the direction of the field modulation for $q
\ge 5$ where the quasiliquid phase emerges.
We explain this behavior in terms of free-energy landscapes on the
two-dimensional magnetization plane.
\end{abstract}

\maketitle

\section{Introduction}
A weak input signal can be amplified by noise. This is called
stochastic resonance (SR) and there has been a vast amount of theoretical
and experimental studies about this phenomenon~\cite{review}.
For a system with a single degree of freedom, SR can be illustrated by
a particle trapped in a double-well potential but constantly hit by random
noise:
The particle inside one minimum moves to the other due to the noise, and this
happens with a characteristic time scale $\tau$ denoted by the relaxation time.
If we apply weak force that oscillates with frequency $f \sim \tau^{-1}$, which
is termed \emph{time-scale matching condition}, the particle can jump over the
potential barrier back and forth in a periodic manner, amplifying the input
force. 

In many practical situations,
the noise is given by thermal contact with a heat bath, and
$\tau$ thus depends on temperature $T$~\cite{SRintheIsing,Analytic_KI}.
For a system with a single degree of freedom, $\tau$ is described by the
simple Kramer rate~\cite{kramer} which diverges exponentially at $T=0$.
As a result, if we consider the response as a function of $T$, the time-scale
matching condition is usually fulfilled at a single point, and multiple
resonance peaks are observable only when the dynamics has certain
symmetry~\cite{vilar}.
For a system with many degrees of the freedom, on the other hand, $\tau$ is
not necessarily explained in that way: If the system undergoes
a continuous phase transition at $T=T_c$, for example, $\tau$ diverges at
this critical point. In other words, $\tau^{-1}$ has a nonzero value over the
whole temperature region except at $T_c$. Therefore, as long as $f$ is low
enough, the matching condition can be satisfied once above $T_c$ and once
below $T_c$, so the resonance will take place twice as $T$ varies from zero to
infinity. 
The prediction of double peaks has been confirmed in various many-body systems
under periodic perturbations, including classical spin
systems~\cite{SRintheIsing,dsr,mf} as well as a quantum-mechanical
case~\cite{quantum}. 
However, most of these systems share one common feature
that they undergo spontaneous symmetry breaking in the absence of external
perturbations. Our question in this study is how the response changes if a
system
possesses the quasi-long-range order without spontaneous symmetry breaking,
and the two-dimensional (2D) $q$-state clock model~\cite{jkkn,*elit} can be the
best candidate to systematically investigate this problem. This model has
played an important role in a 2D melting scenario~\cite{melting}, and some
experimental studies suggest a connection of this model to domain pattern in
ferroelectric materials~\cite{ferro}.

\begin{figure}
\includegraphics[width=0.45\textwidth]{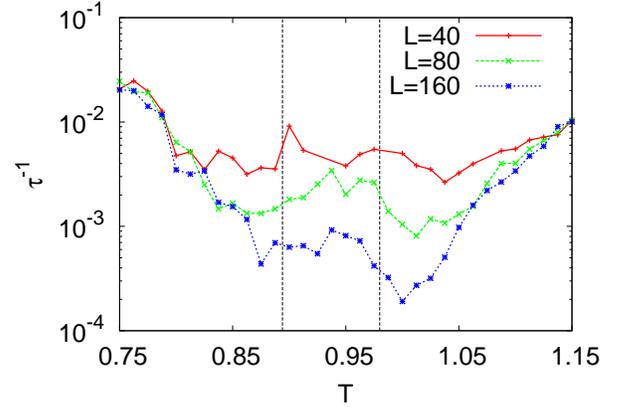}
\caption{(Color online)
Relaxation time $\tau$ of the five-state clock model in equilibrium at
$\mathbf{h}=0$, obtained from the normalized autocorrelation function of $m$
(see Ref.~\cite{avgtau}).
The temperature $T$ is given in units of $J/k_B$, where
$k_B$ is the Boltzmann constant, and
the vertical lines indicate $T_{c1}$ and $T_{c2}$ in the
thermodynamic limit, respectively, estimated in Ref.~\cite{papa1}.
As the system size increases, $\tau^{-1}$ vanishes in the
quasiliquid phase between $T_{c1}$ and $T_{c2}$, where $\mathbf{m} =
me^{i\phi}$ may freely wander around in the angular direction.
}
\label{fig:matching}
\end{figure}

\begin{figure*}
\includegraphics[width=\textwidth]{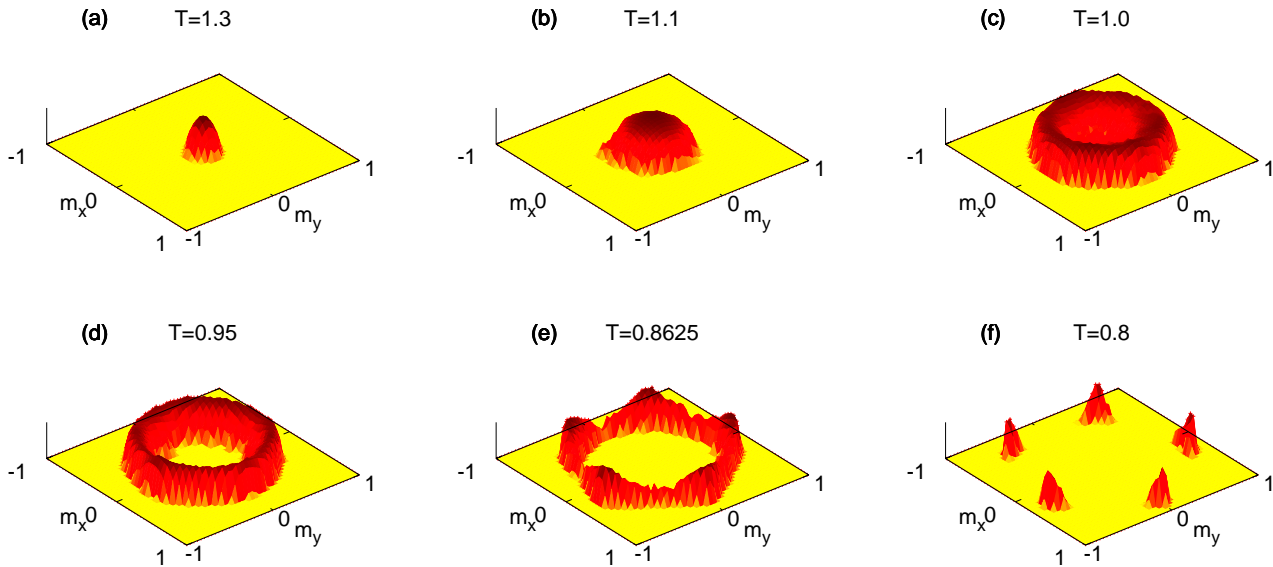}
\caption{(Color online)
Free-energy landscapes on the magnetization plane, drawn upside
down for the five-state clock model with size $N=80 \times 80$ (see text
for details).}
\label{fig:freeE}
\end{figure*}

Let us review some equilibrium properties of this model.
The Hamiltonian of the $q$-state clock model in the $L\times L$ square
lattice is written as
\begin{equation}
H = -J \sum_{\left<j,k\right>} \mathbf{S}_j \cdot \mathbf{S}_k - \mathbf{h}
\cdot \sum_{j} \mathbf{S}_j,
\label{eq:hamiltonian}
\end{equation}
where $J$ is a coupling constant,
$\sum_{\left<j,k\right>}$ runs over the
nearest neighbor pairs and $\mathbf{h}$ is an external magnetic field. Each
spin $\mathbf{S}_j$ at site $j$
has a discrete angle $\theta_j = 2\pi n_j/q$
with $n_j = 0, 1, 2,\ldots, q-1$.
If $q=2$, the model reduces to the Ising model, and
it approaches the $XY$ model as $q\rightarrow \infty$.
The magnetization is given as a 2D vector $\mathbf{m} = N^{-1} \sum_j
\mathbf{S}_j$, where $N \equiv L^2$ is the total number of spins,
and it can also be written as a complex number $me^{i\phi} = N^{-1} \sum_j
e^{i\theta_j}$ with $m \equiv |\mathbf{m}|$.
Suppose that the field $\mathbf{h}$ is absent.
For $q<5$, the system undergoes a single order-disorder transition
and we may well expect double resonance peaks~\cite{SRintheIsing,dsr,mf}.
On the other hand, if $q\geq5$, there appear two infinite-order phase
transitions,
one at $T_{c1}$ and the other at $T_{c2} (> T_{c1})$~\cite{q5,residual}.
In the disordered phase at $T>T_{c2}$, the spins are randomly rotated by
thermal fluctuations to one of the $q$ possible directions so that two spins
are not much correlated if placed just a few lattice spacings apart.
It is obvious that $m$ vanishes in this phase.
When $T<T_{c1}$, on the other hand, almost all the spins
point in the same direction, yielding nonzero $m$. In this ordered phase,
thermal fluctuations are so weak that a spin can only individually deviate
from the preferred direction every once in a while. It implies that there is
no appreciable collective mode and we again find short-ranged correlation in
spin fluctuations.
The intermediate phase between $T_{c1}$ and $T_{c2}$ is actually more
interesting than the other two surrounding it,
because the spin-spin correlation decays algebraically with a diverging
correlation length $\xi \rightarrow \infty$.
The spin relaxation time $\tau$ also diverges because $\tau \sim \xi^z$ with
a dynamic critical exponent $z>0$.
This intermediate phase is sometimes dubbed \emph{quasiliquid} due to the
nontrivial correlations in space and time.
Since $\tau^{-1}$ is zero in
the quasiliquid phase (Fig.~\ref{fig:matching}), we deduce that
the time-scale matching condition can be satisfied below $T_{c1}$ and
above $T_{c2}$, but not in between.

\begin{figure}
\includegraphics[width=0.45\textwidth]{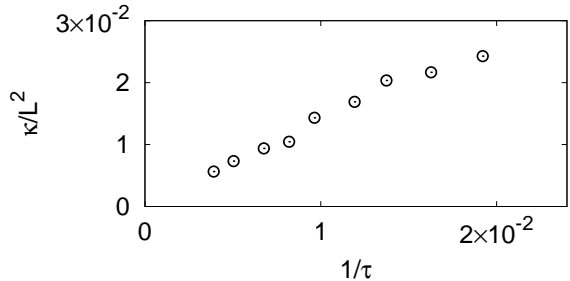}
\caption{Comparison between the inverse relaxation time, measured from the
autocorrelation of $m$, and the curvature of the free-energy landscape (see
Fig.~\ref{fig:freeE}) near its minimum with varying $T$. Each data point
represents a different $T$ in the high-temperature phase ($1.1 \leq T
\leq 1.2$) for the system of the size $L=80$. The error bar is smaller
than the symbol size.
}
\label{fig:compare}
\end{figure}

It is instructive to see the free-energy landscape $f(\mathbf{m})$ in the 2D
magnetization plane (Fig.~\ref{fig:freeE}).
It can be estimated as $f(\mathbf{m}) \propto
-k_B T\ln{p(\mathbf{m})}$, where $k_B$ is the Boltzmann constant and
$p(\mathbf{m})$ means the probability to observe $\mathbf{m}$ in Monte Carlo
(MC) simulations. We have obtained
Fig.~\ref{fig:freeE} by simulating the model on a $80 \times 80$ square lattice,
where both $J$ and $k_B$ are set to unity.
For better visualization, the landscapes are drawn upside down so that
a free-energy minimum appears as a peak.
In the disordered phase at $T>T_{c2}$,
the free-energy landscape has a global minimum at the
center [Fig.~\ref{fig:freeE}(a)], which implies $m=0$ in the
thermodynamic limit as explained above. One should note that the nonzero $m$
in the disordered and the quasiliquid phases is a finite-size effect which
eventually vanishes as $N \rightarrow \infty$.
In the disordered phase,
$\mathbf{m}$ can take any angle $\phi$ between zero and $2\pi$, and the
minimum gets broader as $T$ decreases [Fig.~\ref{fig:freeE}(b)].
It is important that the transition at $T_{c2}$ is not involved with
spontaneous symmetry breaking [Figs.~\ref{fig:freeE}(c) and
\ref{fig:freeE}(d)], and
the breaking happens only when $T$ is lowered further down to $T_{c1}$
[Figs.~\ref{fig:freeE}(e) and \ref{fig:freeE}(f)].
When the system is perturbed slightly from a minimum of this free-energy
landscape $f(\mathbf{m})$, we expect $d\mathbf{m}/dt \propto -\partial f /
\partial \mathbf{m}$~\cite{mf}. In other words, if the minimum is approximated
as $f(\mathbf{m}) \approx \frac{1}{2} \kappa |\mathbf{m}|^2$ in the
high-temperature phase, our guess is that the coefficient $\kappa$ will be
inversely proportional to the relaxation time $\tau$ since $d\mathbf{m}/dt
\propto -\kappa \mathbf{m}$. Such a relation is well substantiated in
Fig.~\ref{fig:compare}. If we further extend this observation to lower
temperatures,
the shapes of the landscapes immediately suggest the existence of two
different time scales, i.e., one in the radial direction and the other in the
angular direction, which we denote by $\tau_\parallel$ and $\tau_\perp$,
respectively. Even if a time-dependent external field is applied, as long as
it is weak enough, the free-energy picture can still provide us with
qualitative understanding.
We therefore expect from Fig.~\ref{fig:freeE} that the SR behavior will
depend on the modulating direction of $\mathbf{h}$ when $q\geq 5$ and $T<T_{c2}$.

This speculation is readily confirmed by our numerical calculations:
By measuring correlation between $\mathbf{m}$ and $\mathbf{h}$
as will be detailed below, we observe the followings:
When a time-varying field $\mathbf{h}$ is applied in the
direction of $\mathbf{m}$, we find that the transition
point $T_{c2}$ is sandwiched between two SR peaks as in the Ising
case~\cite{SRintheIsing,dsr,mf}, though
the one below $T_{c2}$ is broadened all the way down to $T_{c1}$.
If the driving field $\mathbf{h}$ is orthogonal to $\mathbf{m}$, the
behavior becomes radically different because
it is the whole
quasiliquid phase rather than a single point that is surrounded by two
SR peaks.
We first begin with explaining our
numerical method in the next section and then present the results in
Sec.~\ref{sec:Results}. With discussing physical implications of our results,
we conclude this work in Sec.~\ref{sec:Conclusion}.

\section{Numerical method}
\label{sec:Numerical method}
In case of the Ising model ($q=2$), it is a common practice to use the kinetic
Glauber-Ising dynamics~\cite{Glauber} when one studies its behavior slightly out
of equilibrium~\cite{Analytic_KI,dsr}. It is worth noting that this approach has
achieved qualitative agreements with experimental observations such as dynamic
hysteresis~\cite{rmp}.
We need to generalize the Glauber dynamics
for simulating the $q$-state clock model~\cite{mf}, the result of which
essentially corresponds to the heat-bath algorithm among MC
methods~\cite{heatbath,*newman}. Although MC algorithms are not meant to
simulate dynamic properties, it has been widely accepted that they can
effectively describe real dynamics as long as equipped with a local update rule
and a small acceptance ratio~\cite{nowak,*mcdynamics}.
So our algorithm works as follows:
We randomly choose a spin, say, $\mathbf{S}_j$ with $\theta_j = 2\pi n_j /q$,
and calculate how much the energy would change if the angle was switched to
$\theta_j' = 2\pi n_j'/q$. Denoting this amount of change by $\Delta E(n_j')$,
the probability to choose $n_j'$ as its next value is given as $p(n_j') \propto
\exp[-\Delta E(n_j')/T]$ with a normalization condition $\sum_{n_j'} p(n_j') =
1$.

The underlying geometry is the $L \times L$ square lattice with periodic
boundary conditions, the size of which varies between $L=40$ and $160$.
The time $t$ in MC simulations is measured in units of one MC time
step which corresponds to $N$ MC tries for spin update.
We will fix the field amplitude $h_0 = 10^{-2}$ and frequency $f =
10^{-3}$, respectively, throughout this work.
As mentioned above, we may consider two different field directions:
Since the angle of $\mathbf{m}$ is denoted as $\phi$, the field in the
parallel direction is written as $\mathbf{h}_\parallel(t) = h_0 \cos{2\pi f
t}(\cos{\phi}, \sin{\phi})$, while in the perpendicular
direction it is written as $\mathbf{h}_\perp(t)=h_0 \cos{2\pi f t}(-\sin{\phi},
\cos{\phi})$.
Since $\phi$ is also time-dependent, we need to measure it at the beginning
of each period to adjust the field direction, but
it should be kept fixed within the period. 
We are going to apply either $\mathbf{h}_\parallel$ or
$\mathbf{h}_\perp$ to the system and compare the responses. We point
out that an external field ${\mathbf h}$ in a fixed direction can be
decomposed into two components (parallel and perpendicular to
${\mathbf m}$) and the system's response contains contributions from
both components. We drive the system either by $\mathbf{h}_\parallel$ 
or by $\mathbf{h}_\perp$ only to identify physical mechanism of the
resonance behavior more clearly in comparison with the
temperature-dependent free-energy landscape in Fig.~\ref{fig:freeE}.
Our main observable is defined as
\begin{equation}
D \equiv \left \langle \frac{1}{h_0 \Lambda} \int_{n\Lambda}^{ (n+1)\Lambda} \mathbf{m}\cdot
\mathbf{h} ~dt
\right \rangle,
\end{equation}
where the integral is over one period $\Lambda \equiv f^{-1}$ and the bracket
means the average over $n \in [101,900]$ with the transient behavior in early
times ($n \in [0,100]$) neglected.
In one limiting case where
$T \rightarrow \infty$, $D$ should be identically zero since $m$ vanishes there.
In the other
limiting case where $T \rightarrow 0$, $\mathbf{m}$ is frozen regardless of
the small perturbation
$\mathbf{h}$ so that the integral of cosine over one period yields zero
again. Only when $\mathbf{m}$ runs closely after $\mathbf{h}$, the integrand
gives positive contribution on average, and we interpret a large value of
$D$ as signaling the stochastic resonance behavior. At the same time, one
should note that $\mathbf{m} \perp \mathbf{h}$ may also induce vanishingly small
$D$ even if $\mathbf{m}$ does vary in time.

\begin{figure*}
\includegraphics[width=0.8\textwidth]{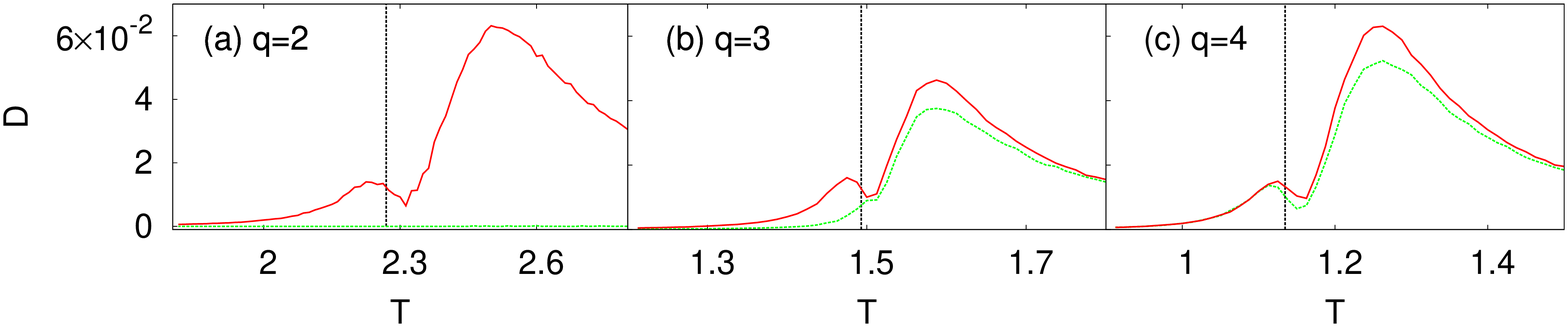}
\caption{(Color online)
Responses for $q<5$, where the solid (red)  and dashed (green) curves
show results under $\mathbf{h}_\parallel$ and $\mathbf{h}_\perp$,
respectively. (a) When $q=2$, the system can respond only to
$\mathbf{h}_\parallel$. (b) When $q=3$ or (c) $q=4$, double SR peaks appear
in any of the field directions. The vertical dotted lines indicate
the equilibrium
critical points in the thermodynamic limit when the field is absent.}
\label{fig:q234}
\end{figure*}
\begin{figure}
\includegraphics[width=0.48\textwidth]{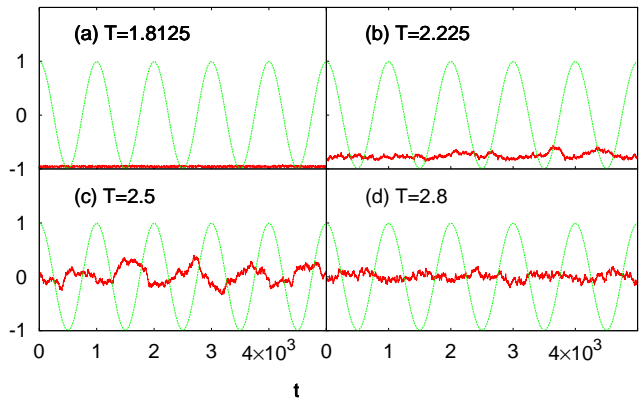}
\caption{(Color online)
The solid  (red) lines represent $m$ for $q=2$, and the sinusoidal
dashed (green) lines show the normalized external field, $\cos 2\pi ft$. (a) When $T$ is
low, the system is frozen at a symmetry-broken state. (b) As $T$ increases,
$m$ begins to fluctuate but the amplitude is small because it is trapped in
a narrow free-energy minimum. (c) The strongest response is found around the
resonance peak above $T_c$ where $m(t)$ moves between the positive and
negative sides. (d) If $T$ increases further, the free-energy minimum at the
origin ($m=0$) gets narrow and the resonance is thus suppressed.}
\label{fig:time_m}
\end{figure}

\begin{figure*}
\includegraphics[width=0.8\textwidth]{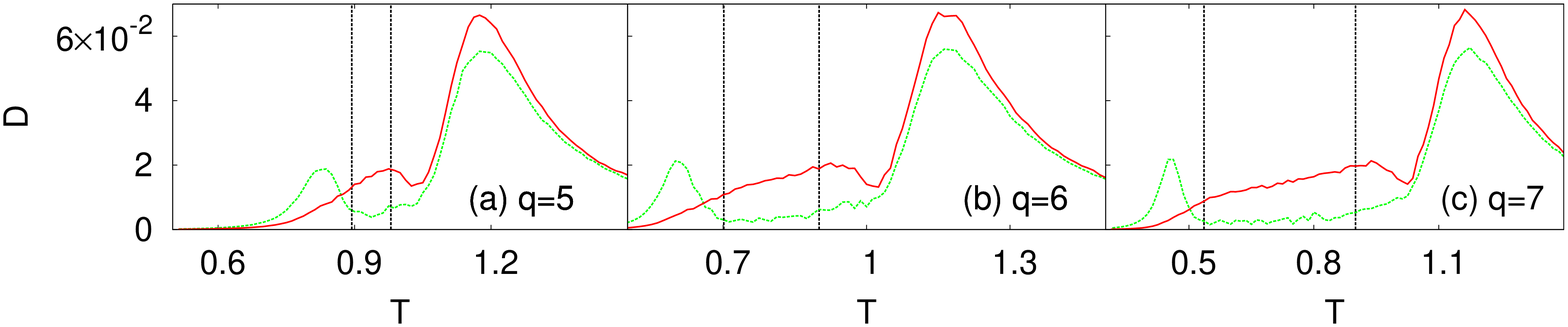}
\caption{(Color online)
Double SR peaks for $q\ge 5$, where the solid (red) and dashed (green)
lines mean the results under $\mathbf{h}_\parallel$ and $\mathbf{h}_\perp$,
respectively. The equilibrium critical points in the thermodynamic limit are
represented as the vertical dotted lines.
While strong responses are found below
and above $T_{c2}$ under $\mathbf{h}_\parallel$, the quasiliquid phase
shows little $D$ under $\mathbf{h}_\perp$ so the left peak is located
below $T_{c1}$. Note that $T_{c1}$ ($T_{c2}$) tends to be underestimated
(overestimated) for finite $N$. }
\label{fig:q567}
\end{figure*}

\section{Results}
\label{sec:Results}

In this section, we present MC results obtained only for $L=80$, because
the qualitative features remain unaltered for larger systems and there is
little size dependence in peak heights as well.

If $q<5$, the system undergoes a single continuous phase transition.
Therefore, the observable $D$ shows the expected double-peak structure,
one below and the other above $T_c$ when $\mathbf{h}_\parallel$ is applied
(Fig.~\ref{fig:q234}). Even though $\mathbf{h}_\perp$ has no physical meaning
in the Ising case ($q=2$), it induces qualitatively the same responses as
$\mathbf{h}_\parallel$ does when $q=3$ or $q=4$.
It is notable that the two peaks are highly asymmetric in each plot,
which means that the
system amplifies the signal better at the second peak above $T_c$
(Fig.~\ref{fig:time_m}). This asymmetry is characteristic of a
low-dimensional system, in contrast to the mean-field (MF) case~\cite{mf}:
In the MF case with a given field frequency $f$, the
response is fully specified by $\tau$. Since each of the double
peaks is characterized by the same condition that $\tau \sim f^{-1}$, the
peak height is accordingly the same as well. Returning back to the 2D
case, we see that the asymmetry is actually plausible because the system is
more susceptible in the disordered phase. It is well-known that the static
susceptibility $\chi$ around $T_c$ behaves as $\chi_{\pm} \sim \Gamma_{\pm}
|T-T_c|^{-\gamma}$, where the subscript means the sign of the reduced
temperature $(T-T_c)/T_c$, and $\gamma$ is a critical exponent of the model.
A prediction from the renormalization-group theory is that
the amplitude ratio between $\Gamma_+$ and $\Gamma_-$ is universal, whereas
they are not individually.
The universal amplitude ratio is exactly calculated as $\Gamma_+/\Gamma_- =
37.6936520\ldots$ for $q=2$ and $q=4$~\cite{baro,*delfino}, and estimated as
$\Gamma_+/\Gamma_- \approx 13.86(12)$ for $q=3$~\cite{shchur}.
It is reasonable to guess that a relevant factor
to the peak height will be $\left<\Delta m \right> = \sqrt{\left< m^2
\right> - \left<m\right>^2} \propto \chi^{1/2}$. In other words, our guess
is that the ratio between the peak heights $D^\ast_+/D^\ast_-$ is roughly
proportional to $(\chi_+/\chi_-)^{1/2} \sim (\Gamma_+/\Gamma_-)^{1/2}$ so
that $\zeta \equiv (D^\ast_+/D^\ast_-)^{-1}(\Gamma_+/\Gamma_-)^{1/2}$ yields
similar values when $q$ varies between $2$ and $4$. Although this
argument is not meant to be exact and the estimates of $D^\ast_+/D^\ast_-$
are not precise either, this explains some part of the observation because
we indeed find $\zeta =
1.32(12)$, $1.31(7)$, and $1.44(4)$ for $q=2, 3$, and $4$, respectively.
Moreover, these values are comparable to the MF result
$\zeta_{\rm MF} = \sqrt{2} = 1.414\ldots$ since we already know
$D^\ast_+/D^\ast_-=1$ and the Landau theory predicts $\Gamma_+/\Gamma_- = 2$.

It is straightforward to perform the same simulations for $q\ge 5$, but the
behavior is rather different depending on the field direction as expected
(Fig.~\ref{fig:q567}).
The response is insensitive to the direction in the disordered
phase since $\mathbf{m}$ has no
meaningful direction with vanishingly small magnitude. Below $T_{c2}$,
however, the dependence on the field direction is clearly visible, which can be
understood by using the free energy landscape (Fig.~\ref{fig:freeE}),
provided that the field is so weak that the system remains close to
equilibrium. According to this picture,
in the quasiliquid phase, there is no significant free-energy barrier in the
angular direction: This implies very large $\tau_\perp$,
whereas $\tau_\parallel$ remains always finite because the system is
effectively confined in a free-energy well in the radial direction. This
explains why the SR peak is observed only under $\mathbf{h}_\parallel$ in
this phase.
It is below $T_{c1}$ that the system experiences
free-energy barriers in the angular direction. This barrier
regulates the divergence of $\tau_\perp$, and a clear
resonance peak is thereby developed under $\mathbf{h}_\perp$.
For an arbitrary field direction, the response of the system is
described as a combination of the results under $\mathbf{h}_\parallel$ and
$\mathbf{h}_\perp$, because we are working in the linear-response regime.
We have also measured peak height ratios when $q \ge 5$ for the sake of
completeness:
Under $\mathbf{h}_\parallel$, we estimate $D^\ast_+/D^\ast_-$ as $3.55(11)$,
$3.4(2)$, and $3.2(1)$ for $q=5, 6$, and $7$, respectively. If we apply
$\mathbf{h}_\perp$ instead, the estimates of $D^\ast_+/D^\ast_-$ now read
$2.91(7)$, $2.6(1)$, and $2.5(1)$, respectively. It is interesting that
$q=6$ and $7$ are so similar in this respect that the values in either
direction are on top of each other within the errorbars.

\section{Conclusion}
\label{sec:Conclusion}

\begin{figure}
\includegraphics[width=0.2\textwidth]{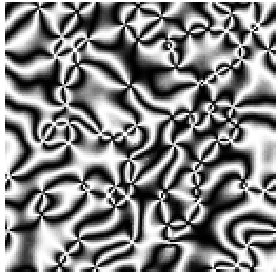}
\caption{
Spin configuration of the 2D clock model in the large-$q$ limit quenched from $T=\infty$ to
$T=0$. Each spin direction $\theta_j$ is expressed as brightness
proportional to $\sin^2 2\theta_j$.}
\label{fig:sch}
\end{figure}

We have investigated responses of the 2D $q$-state clock system under external
oscillating fields.
Double resonance peaks are found below and above the
unique critical point $T_c$
for $q<5$, and the peak positions are not sensitive to the field direction.
For $q\ge 5$, however, the emergence of the quasiliquid phase in 2D
makes the situation more complicated than the MF analysis in that
the resonance behavior crucially depends on the field
direction, especially when $T < T_{c2}$. We have qualitatively explained
this difference by using the free energy landscape. Of course, the
free-energy argument implies that we have restricted ourselves to the
linear-response regime, which loses validity as the applied field
becomes stronger.

We may also interpret the directional dependence in the quasiliquid phase
in the context of liquid crystal (LC)~\cite{degennes}:
Suppose that each LC molecule carries a small electric dipole moment and can
be described as an $XY$-typed spin variable.
If a thin LC film is exposed to linearly polarized light,
the oscillating electric field interacts with each electric dipole,
and the periodically driven dipole in turn emits
electromagnetic waves as a response.
Our observation in this work suggests how the response
will depend on the relative orientation between the dipole moment and
the polarization of the incident light: If they are perpendicular to each
other, for example, the molecule will respond to the input with a large phase
delay due to the continuous symmetry, as indicated by small $D$. As a
consequence, secondary wave will interfere destructively with the
primary one. 
In addition, when LC is placed between two crossed polarizers,
the so-called Schlieren texture~\cite{degennes} captures spatial variations in
orientations of the LC molecules.
In a simple thought experiment where these polarizers are taken into account,
as the direction
$\phi$ of LC rotates from zero to $2\pi$,
the final intensity of light through the second polarizer will have
four minima at $\phi = 0, \pi/2, \pi$, and $3\pi/2$.
By numerically simulating the quasiliquid phase of a 2D clock model with very
large $q$, we illustrate a possible optical image in Fig.~\ref{fig:sch},
which precisely reproduces a typical Schlieren texture in real experiments.

\acknowledgments

This work was supported by the National Research
Foundation of Korea (NRF) grant funded by the Korea government (MEST) (Grant
No. 2011-0015731).

\bibliographystyle{apsrev4-1}
\bibliography{dsr}

\end{document}